\documentclass[12pt,preprint]{aastex}

\newcommand{\rstar}{\,\mbox{$\mbox{R}_*$}}
\newcommand{\degrees}{\mbox{$^\circ$}}
\newcommand{\kmsec}{\,\mbox{$\mbox{km}\,\mbox{s}^{-1}$}}
\newcommand{\vsini}{\mbox{$v_e\,\sin\,i$}}
\newcommand{\prot}{\mbox{$P_{\mbox{rot}}$}}

\begin{document}
\title{Inferring coronal structure from X-ray lightcurves and  Doppler shifts: 
a {\em Chandra} study of AB Doradus.}
\author{G.A.J. Hussain\altaffilmark{1,2}, 
N.S. Brickhouse\altaffilmark{2}, 
A.K. Dupree\altaffilmark{2}, 
M.M. Jardine\altaffilmark{3}, 
A.A. van Ballegooijen\altaffilmark{2},  R. Hoogerwerf\altaffilmark{2},
A. Collier Cameron\altaffilmark{3},
J.-F. Donati\altaffilmark{4} 
and F. Favata\altaffilmark{1}} 
\altaffiltext{1}{Astrophysics Division, 
Research \& Science Support, 
Department of ESA, ESTEC, Postbus 299, 
Noordwijk, The Netherlands}
\altaffiltext{2}{Harvard-Smithsonian Center for Astrophysics,  
60 Garden Street, Cambridge MA 02138, USA}
\altaffiltext{3}{School of Physics \& Astronomy,  
University of St Andrews, North Haugh, Fife KY16 9SS, UK}
\altaffiltext{4}{Laboratoire d'Astrophysique,  
Observatoire Midi-Pyr\'en\'ees, Avenue E. Belin, F-31400, Toulouse, France}

\begin{abstract}
The {\em Chandra} X-ray observatory monitored the single cool star, 
AB Doradus, continuously for a period lasting 88\,ksec 
(1.98\,$P_{\rm rot}$)  in 2002 December with the LETG/HRC-S. 
The X-ray lightcurve shows rotational modulation, 
with three peaks that repeat  in two consecutive rotation cycles. 
These peaks may indicate the presence of  compact emitting regions 
in the {\em quiescent} corona.
Centroid shifts as a function of phase in the strongest line
profile, O\,{\sc viii}\,18.97 \AA, indicate Doppler rotational velocities with
a semi-amplitude of $30 \pm 10$\kmsec. 
By taking these diagnostics into account along with constraints
on the rotational broadening of line profiles 
(provided by archival  Chandra HETG \ion{Fe}{17}
and FUSE \ion{Fe}{18} profile) we can construct a simple model of the X-ray 
corona that requires two components. One of these components is responsible for 
80\% of the X-ray emission, and arises  
from the pole and/or a homogeneously distributed corona.  
The second component consists of two or three compact active regions
that cause  modulation in the lightcurve and  contribute to the
O\,{\sc viii} centroid shifts.
These  compact regions account for 16\% of the 
emission and are located near the stellar surface 
with heights of less than 0.3\rstar. 
At least one of the compact active regions is located 
in the partially obscured hemisphere of the inclined star, while 
one of the other active regions may be located at $40$\degrees.
High quality X-ray data
such as these can test the models of the coronal magnetic field
configuration as inferred from magnetic Zeeman Doppler imaging.
\end{abstract}
\section{Introduction}
Stellar coronae affect how ZAMS stars spin down,
how binary systems interact, and even how planets form.
Magnetically driven winds control the spin-down rates
of young main sequence low mass stars; the latitudes
at which these winds originate affect the time-scale on which the star 
will spin down (Solanki, Motamen \& Keppens 1997).
Heating from active coronae also affects planetary atmospheres, 
e.g. XUV radiation by young
active stars significantly affects atmospheric escape rates for surrounding
giant exo-planets, potentially causing them  to evaporate entirely at close
orbital distances (Lammer et al. 2003).

X-ray and EUV observations from the {\em Chandra}, 
XMM-{\em Newton} and EUVE satellites have revealed that 
coronae in active cool stars are unlike anything 
observed on the Sun (Dupree et al. 1993). Emission measure
distributions (EMDs) in even relatively quiescent coronae 
are different from the Sun, with higher levels of emission 
($10^{52}$\,cm$^{-3}$) and a 
markedly different  shape (with a sharp peak at 8\,MK).
Furthermore, electron densities are greater than $10^{10}$\,cm$^{-3}$ in 
plasma at temperatures of 8\,MK and beyond 
(e.g. {G\"udel} et al. 2001; Sanz-Forcada et al. 2002). 
To compare, in active regions on the Sun, emission measures
peak at about $10^{50}$\,cm$^{-3}$ at 2.5\,MK 
(Brickhouse, Raymond \& Smith 1995;  Orlando, Peres \& Reale  2000). 

While the thermal properties of active 
stellar coronae are increasingly well-determined, we 
have yet to establish where the emitting plasma is located
relative to the stellar surface.
It is difficult to build a detailed picture of coronal 
structure as active stars are too distant to resolve 
 at X-ray and EUV wavelengths. We therefore have 
to rely on indirect techniques, only rarely applicable, to 
evaluate the location and extent of the emitting corona 
(see review by Schmitt 1998).
For example, periodic photometric variations in the EUV flux of the contact
binary 44i~Boo strongly constrained the location of the emission
in a small, dense high-latitude region on the primary star 
(Brickhouse \& Dupree 1998);
the long exposure time (covering 19 orbital periods) and relatively low levels
of flaring facilitated these measurements. 
On the other hand, large X-ray flares can dominate the emission, 
and thus eclipse mapping of  flares can also reveal the location of X-ray plasma as the
flares originate in localized regions. Flaring loop lengths
 can be estimated from rise-times and decay time-scales in X-ray lightcurves,
assuming a single heating event (Favata et al.  2000). 
Positions can be determined if the flare is eclipsed by the companion
in an eclipsing binary system, 
e.g.  a flare observed in the eclipsing binary, Algol (B8V+K2IV), 
was completely eclipsed,  thus strongly constraining its location 
to 0.5\,\rstar\ above the ``south'' pole of the K star 
 (Schmitt \& Favata 1999; Favata \& Schmitt 1999). 
Modelling of X-ray flares in the eclipsing binary, YY Gem (dM1e+dM1e),
also indicates the presence of compact flaring loops around
one of its components (G\"udel et al. 2001, Stelzer et al. 2002).

Spectra from the {\em Chandra} and XMM-{\em Newton} telescopes are of 
sufficiently high resolution ($100 < R < 2500$)  that individual 
spectral line profiles can now be  analyzed. 
Brickhouse et al. (2001) exploited the spectral resolution 
in a {\em Chandra}/HETG dataset of the contact binary,
44i~Boo (G2V+G2V) to evaluate the location of the emitting 
corona. They measured Doppler shifts as a function of orbital phase in the centroids of 
emission line profiles and found strong evidence for rotational 
modulation. Taking this into account along with the modulation observed in the 
X-ray lightcurve of the dataset, they find that the regions causing 
the velocity variations are located near the pole of the primary star. 

AB Doradus (AB Dor, HD36705) is a well-studied example of a single, active K0 dwarf 
that has recently arrived on the main sequence. 
It rotates very rapidly ($P_{\rm rot}=0.51479$\,day,  \vsini=90\kmsec) 
so two rotation cycles can be covered in just over a day. 
AB Dor also shows signs of magnetic activity at almost 
all wavelengths; it shows strong photometric variability due to dark 
surface spots  and is very X-ray luminous,
$L_{\rm X}/L_{\rm bol}\approx10^{-3}$ (Vilhu \& Linsky 1987). 
Its emission measure distribution (EMD) derived using EUVE 
data is typical for an  active cool star, whether an RSCVn binary
or a ZAMS star
(Sanz-Forcada et al. 2002; Sanz-Forcada et al. 2003). 
Surface spot maps of AB Dor also show activity
patterns typical of active cool stars; with a dark cool polar spot, 
extending down to about 70\degrees\ latitude,  
 co-existing with lower latitude spots 
(e.g. Donati et al. 1999; Unruh, Collier Cameron \& Cutispoto  1997). 
The polar spot is extremely stable with a 
lifetime of several years,  while the lower latitude spots 
appear to have periods of about one month (Hussain 2002).

An increasing amount of evidence supports the picture of a 
compact ($<1\,R_*$), hot ($>1$\,MK) corona in AB Dor: 
(a)  density-sensitive Fe-line ratio measurements from EUVE spectra 
suggest that the stable 8\,MK emission peak in  
AB Dor's EMD  is associated with  high densities, $n_e \approx 
10^{12}$ cm$^{-3}$  that are likely to be 
associated with strong magnetic fields near its surface 
(Sanz-Forcada, Brickhouse \& Dupree 2002);
(b) X-ray and UV lightcurves of AB Dor indicate rotational modulation at the 
5--13\% level ({K\"urster} et al. 1997; Brandt et al. 2001),
suggesting that a component of the corona is inhomogeneously 
distributed --  either very near the surface or else concentrated 
in compact, mid-to-low latitude regions in the extended corona; 
(c) a flare observed in AB Dor by {\em Beppo}SAX was found
to be located $0.3\,R_*$ above the surface 
in the circumpolar region of the star (Maggio et al. 2000).
However,  there is also evidence of cooler ($8 \times 10^{3}<T<10^{5}$\,K) 
magnetically-confined material out to distances between 2--6\rstar\ in several 
active cool stars including AB Dor and V711 Tau
(e.g. Collier Cameron, Jardine \& Donati 2002; Walter 2004). 
Transition region UV line widths (e.g. C\,{\sc iii} 1176\,\AA\ and 
O\,{\sc vi} 1032\,\AA) may be 
significantly large due to rotational broadening (Redfield et al. 2002).
It is unclear whether or not this circumstellar material
is also associated with hot ($>$\,MK) coronal plasma. 

We have acquired spectra of AB Dor with the {\em Chandra} low energy transmission
grating (LETG) covering almost two full 
rotation cycles. We use Gaussian-fitting routines to measure 
Doppler shifts in AB Dor's corona and thus to 
evaluate the locations of X-ray emitting regions. Details of 
the {\em Chandra} observations and data reduction procedures are 
outlined in Section 2.
In Section 3 we discuss the results from our analysis of the X-ray lightcurve.
The LETG spectra are  described and analyzed in 
Section 4. In Section 5 we look for evidence of rotational broadening in LETG and 
high energy transmission grating (HETG) 
datasets of AB Dor. Finally, we summarize our main results 
and use them to derive a simple model of the X-ray emitting corona in Sections 6 and 7.

\section{Observations and data reduction}

\begin{figure}
\epsscale{1}
\plotone{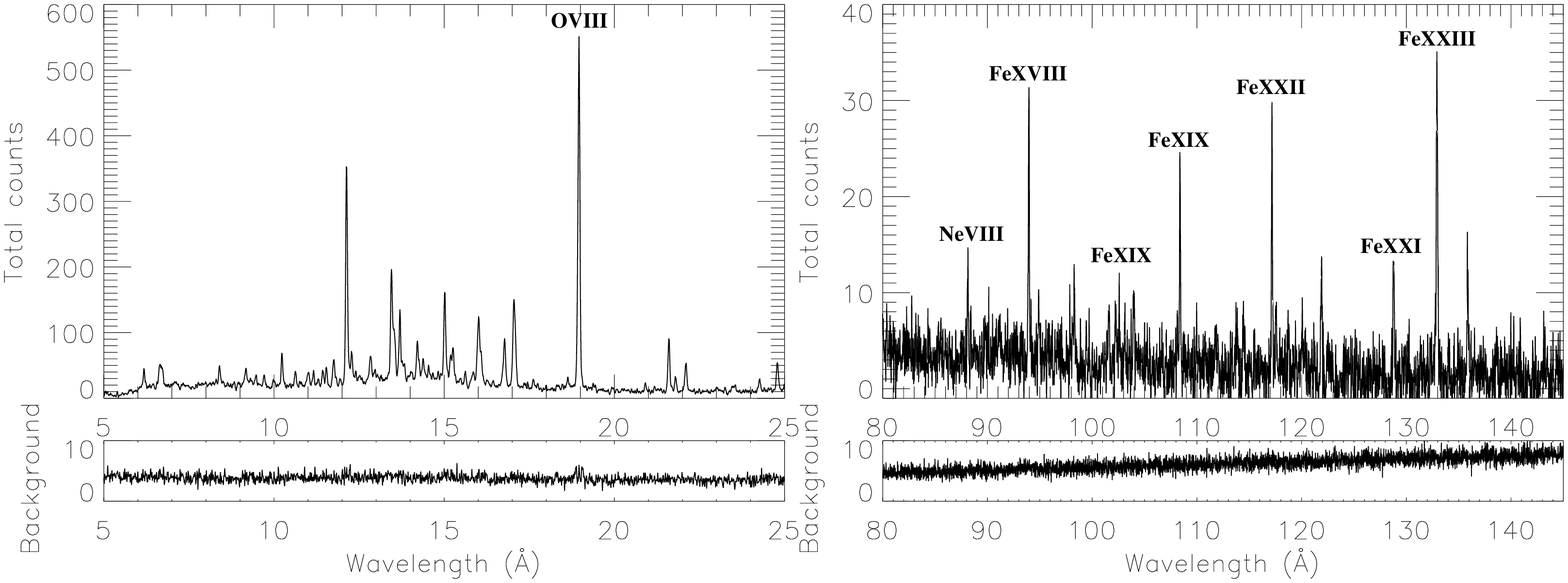}
\caption
{ {\em Chandra} LETG/HRC-S spectrum integrated over 88.1\,ksec (1.98$P_{\rm rot}$). 
The upper panels show the background-substracted spectra in the 
short and long wavelength regions.
The lower panels show the corresponding background levels. Note that
while the spectral resolution in the long wavelength region is better,
this region also has fewer counts and a relatively high background level.
The lines used in this paper are identified in the figure.}
\end{figure}

We observed AB Dor using the {\em Chandra} X-ray Observatory for
88.1\,ksec (1.98\,$P_{\rm rot}$) during one uninterrupted pointing.
This dataset (ObsID 3762) was acquired between
UT 21:25 10 December  and 21:53 11 December 2002
using the LETG with the high resolution camera spectroscopic array (HRC-S). 
This configuration covers the wavelength range, 
$1.2< \lambda < 170 $\,\AA, 
with spectral resolutions ranging from 200 at short wavelengths 
to $\approx$\,3000 at the longest wavelengths 
(see the CXC Proposer's Observatory Guide).

Our aim is to probe the structure of AB Dor's emitting corona by 
searching for rotationally-modulated Doppler shifts in emission line
profiles. X-ray spectra are extracted using the
dedicated data reduction package CIAO v.3.0.1 (Figure\,1).
The extraction procedure followed is standard in all respects except
that the pixel randomization is disabled in order to optimize the
spectral resolution.

\section{Lightcurve analysis}

\begin{figure}
\epsscale{1}
\plotone{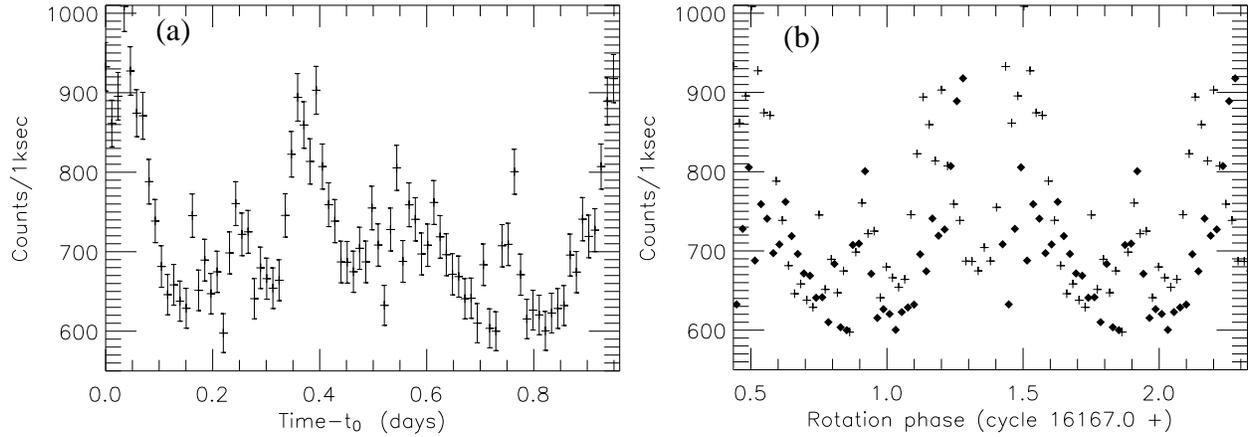}
\caption{ {\bf (a)}~X-ray lightcurve obtained by summing up counts over the low 
background  region of the spectrum ($\lambda < 50$\,\AA). 
{\bf (b)}~X-ray lightcurve folded with AB Dor's rotation period. 
Crosses and diamonds represent consecutive rotation cycles.
The rotation cycle number and phases are  calculated using the  ephemeris 
$H\!J\!D = 2\,444\,296.575 + 0.51479E$  (Innis et al.  1988). }
\label{fig:lightcurve}
\end{figure}

Three sets of lightcurves are extracted to ensure effective subtraction of the
background level: the first of these is the zero-order lightcurve; 
the second lightcurve is derived by integrating counts in the background-subtracted 
short wavelength region ($1 < \lambda < 50$\,\AA) summing up both the
$+1$ and $-1$ orders; and the third lightcurve  is computed by integrating the 
background-subtracted spectra
over all wavelengths in the $+1$ and $-1$ orders. 
All three lightcurves are computed using 1\,ksec bins
and show a high level of agreement, with peaks and plateaus occurring at
the same phases.  We use the second lightcurve in our analysis as it has more counts than 
the zero order lightcurve and a lower background contribution than the third lightcurve 
(which is integrated over the entire spectrum). 
The lightcurve used is shown in Figure\,\ref{fig:lightcurve}a and is  folded on 
AB Dor's rotation period  in Figure\,\ref{fig:lightcurve}b. 
This plot shows a high level of agreement between the 
consecutive rotation cycles, as discussed below. In both rotation cycles, peaks
are observed near phases 0.2, 0.5 and 0.9.

\begin{figure}
\epsscale{1}
\plotone{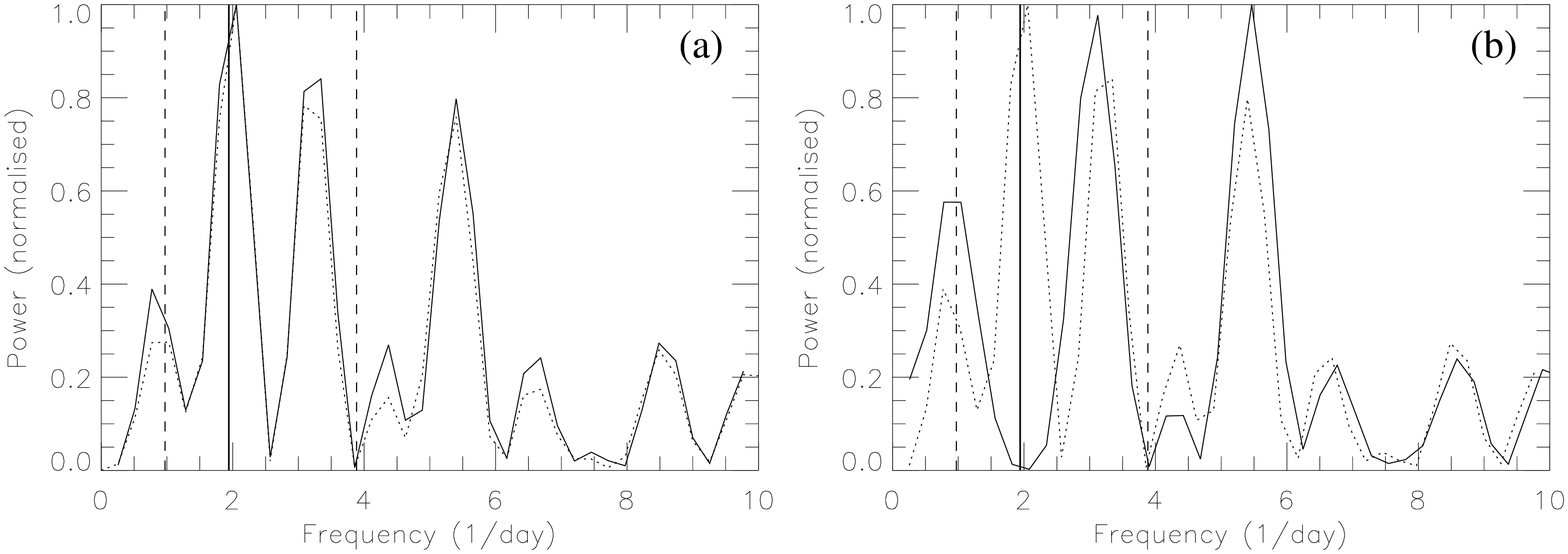}
\caption{ Power spectra for LETG lightcurves shown in Figure\,2.
The power scale is normalised using the maximum power.
The vertical lines mark 2\,$P_{\rm rot}$, 1\,$P_{\rm rot}$ and 0.5\,$P_{\rm rot}$. 
{\bf (a)}~The solid and dotted lines denote power spectra derived using the  
Lomb-Scargle and {\sc clean} algorithms respectively. 
{\bf (b)}~The solid line denotes the power spectrum derived using the Lomb-Scargle
algorithm after subtracting the best-fit sine wave with AB Dor's rotation period
from the X-ray lightcurve. 
The dotted line is the original power spectrum to better enable comparison between 
the two sets of power spectra.  
Note that the peak corresponding to 1\,$P_{\rm rot}$ has been removed. }
\label{fig:period}
\end{figure}

In order to quantify how much 
of the variability evident in this X-ray lightcurve is due to rotational modulation, 
we follow the approach of K\"urster et al. (1997).
The UK {\sc starlink} software package, {\sc period} is used to compute 
the periodograms presented here.
Figure\,\ref{fig:period}a shows that both the Lomb-Scargle and the {\sc clean} algorithms 
produce very similar periodograms. The power scales for the periodograms have been normalised using their peak power for ease of comparison.
With a confidence level of over 95\%, 
the strongest peak is at 2.04\,day$^{-1}$ (0.49 $\pm 0.06$\,day). This is 
consistent with the 0.51\,day rotation period of AB Dor
(marked by a solid vertical line in Figure\,\ref{fig:period}a) 
given the width of this peak.

In order to compute how much spectral leakage may occur from the rotation
period to the other peaks in the periodogram, we fit a sine curve with
AB Dor's rotation period to the observed lightcurve. 
The best-fit sine curve to the dataset has a semi-amplitude of 84\,counts.
As a fraction of the mean amplitude of the LETG lightcurve this corresponds
to rotational modulation at the 12\% level.
This is consistent with  previous estimates of between 5-13\% from
1991 October and November  {\sc rosat} observations of AB Dor  lasting 46\,ksec
in total and spanning 6\,days  (K\"urster et al. 1997).
On subtracting the best-fit sine wave 
from our lightcurve, the period analysis is redone on  this
``cleaned'' lightcurve using the Lomb-Scargle algorithm.
As Figure\,\ref{fig:period}b shows, the strongest peak near the rotation period is removed but the
positions of the other peaks (at 0.32\,day and 0.18\,day) are not significantly affected.

If we look more closely at the phase-folded lightcurve (Figure\,\ref{fig:lightcurve}b) 
we find it can be crudely represented as having a
flat level of activity (with approximately 600 counts
per ksec bin), superimposed with about three peaks.
These peaks in the lightcurves 
repeat fairly consistently in the second rotation cycle
suggesting that they are rotationally modulated and originate in inhomogeneously
distributed, relatively stable, X-ray bright regions.
The degree of repeatability is different for the three peaks. The peak near
phase 0.9 is reproduced extremely well in both phase position (i.e. longitude) and 
peak count rate, indicating a fairly stable structure. The peak near phase 0.2
has the same peak count rate in both cycles, but the peak occurs
about 0.12 later in phase in the second rotation cycle. If the rise
to the peak occurs because of a flaring active region, it seems odd that the 
peak count rate would be so similar. However, as this is near the point at which the observation
ends, we cannot be sure that the rise time is complete. The third peak near phase 0.5
has a much larger count rate in the first cycle, although the phases near the peak count
rates are similar in the two rotation cycles.
 
As the centers of the three lightcurve peaks are 
evenly spaced in phase (phase gaps ranging from 
0.31 to 0.37 ), they are likely the cause of the secondary and 
tertiary peaks in the periodograms in Figure\,\ref{fig:period} which are 
at nearly three times the rotational frequency 5.82\,day$^{-1}$ 
(i.e. 1/3\,\prot) and at half this value, 2.91\,day$^{-1}$.  
K\"urster et al. (1997) have also previously noted short-term periodicities in 
AB Dor's corona from periodogram analyses of X-ray data at previous epochs of 
observation. These peaks in the power spectra probably indicate the numbers of 
peaks and  dips in the lightcurves, caused either by flaring events or from more stable 
rotationally modulated features such as those found here.

\section{Spectral analysis: measuring centroid  shifts}

\begin{figure}
\epsscale{1.0}
\plotone{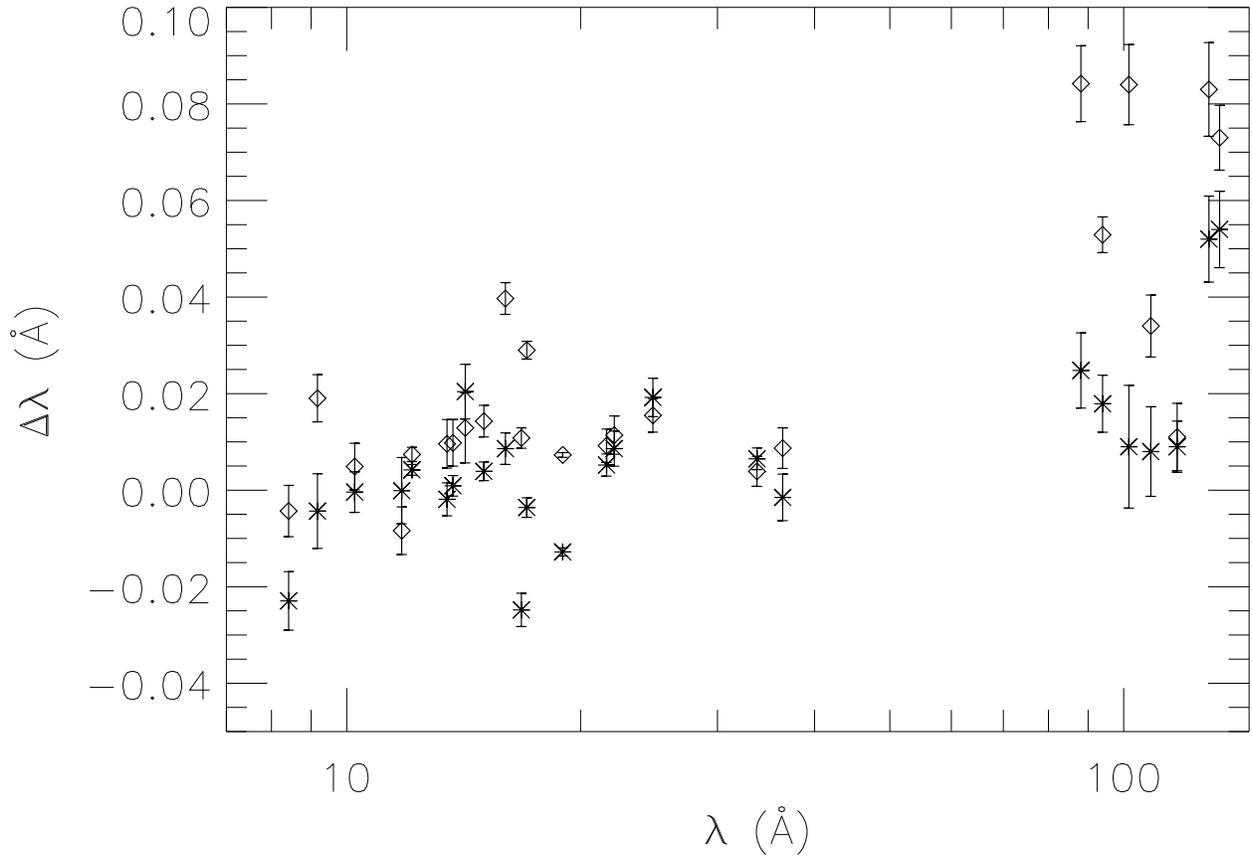}
\caption
{$\Delta\lambda$ is the offset between the observed wavelength position
(measured using the {\em Sherpa} data analysis package) 
and the laboratory position of the line profile (from  the {\sc atomdb}
database). Asterisks represent order $+1$ while diamonds represent order $-1$.}
\label{fig:letgoff}
\end{figure}

We wish to use centroid shifts in X-ray line profiles to evaluate the location of the
emitting corona in AB Dor. Similar methods have been used 
successfully to trace the location of X-ray emitting regions around
binary systems  (Brickhouse et al. 2001; Hoogerwerf, Brickhouse \& Mauche 2004).
However,  we will use this method here to trace the 
locations of the emitting corona in a  {\em single} star for the first time.
When measuring centroid shifts in LETG spectral line profiles in general there are two main 
main challenges: the LETG wavelength scale is not well-determined
and suffers from non-linear deviations from the
laboratory positions of the wavelengths 
(especially at wavelengths greater than 50\AA\ ; Figure\,\ref{fig:letgoff}); 
and secondly, there is some evidence that further deviations can occur on the
the dither time-scale of 1087 seconds (Chung et al. 2004).

We extract a spectrum over the entire 88.1\,ksec exposure and
measure the centroids of the strongest 
line profiles using the Gaussian-fitting 
routine in the data analysis package, {\em Sherpa}.
The absolute wavelength scale is not important as we are looking 
for {\em relative shifts} in the positions of  line centroids as a 
function of phase. 
Offsets between the centroid positions of the line profiles 
and the laboratory wavelengths (using the {\sc atomdb} v.1.2 database)
are computed (Figure\,\ref{fig:letgoff}) and used to recalibrate the 
``zero-velocity'' positions of each line profile. 

The total exposure is divided into eight quarter rotation-phase bins
and spectra are extracted for each of these bins. Note that all 
the bins are of 11.12~ksec length except for the last phase bin which is 
slightly shorter, 10.26~ksec (as the total observation
does not cover two full rotation cycles). 
Because the dither time-scale of the spacecraft is 1087\,sec, 
 each phase bin consists of approximately ten dither cycles. 
Thus the centroid measurements should not be susceptible to wavelength
deviations associated with dithering.

\begin{figure}
\epsscale{1}
\plotone{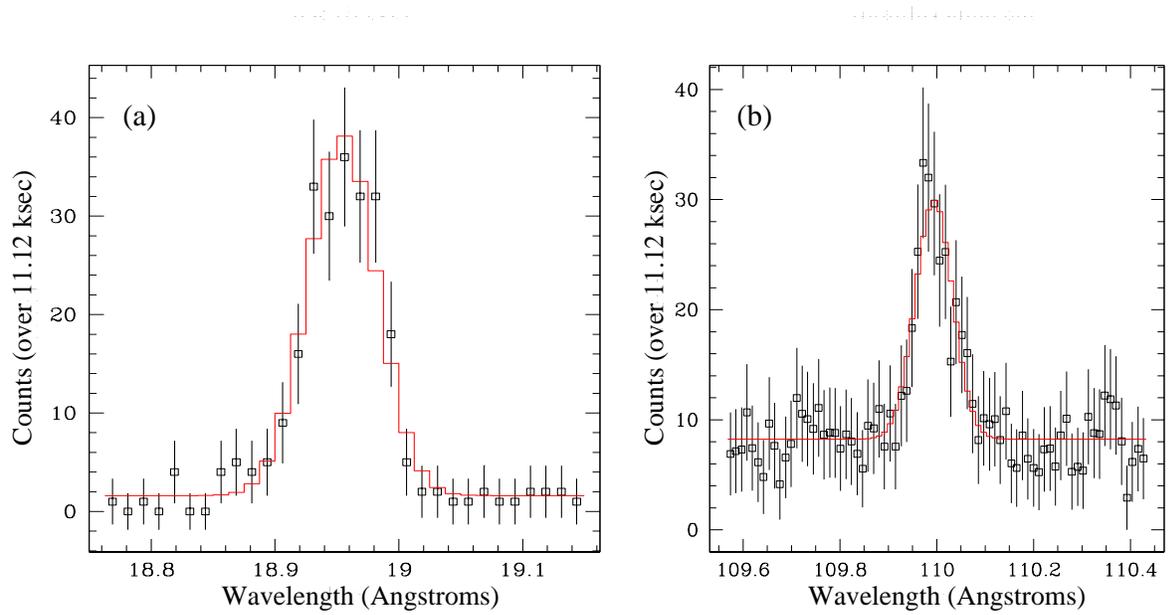}
\caption{ Centroid fits to quarter-phase binned line profiles.
The squares are the data and the solid lines are the Gaussian fits to the data
obtained using the {\em Sherpa} package.
{\bf (a)}~The O\,{\sc viii} 18.97\AA\ line profile.
{\bf (b)}~The profile obtained by summing up the seven 
strongest lines from the long wavelength region.}
\label{fig:letglins}
\end{figure}

\begin{figure}
\epsscale{1.0}
\plotone{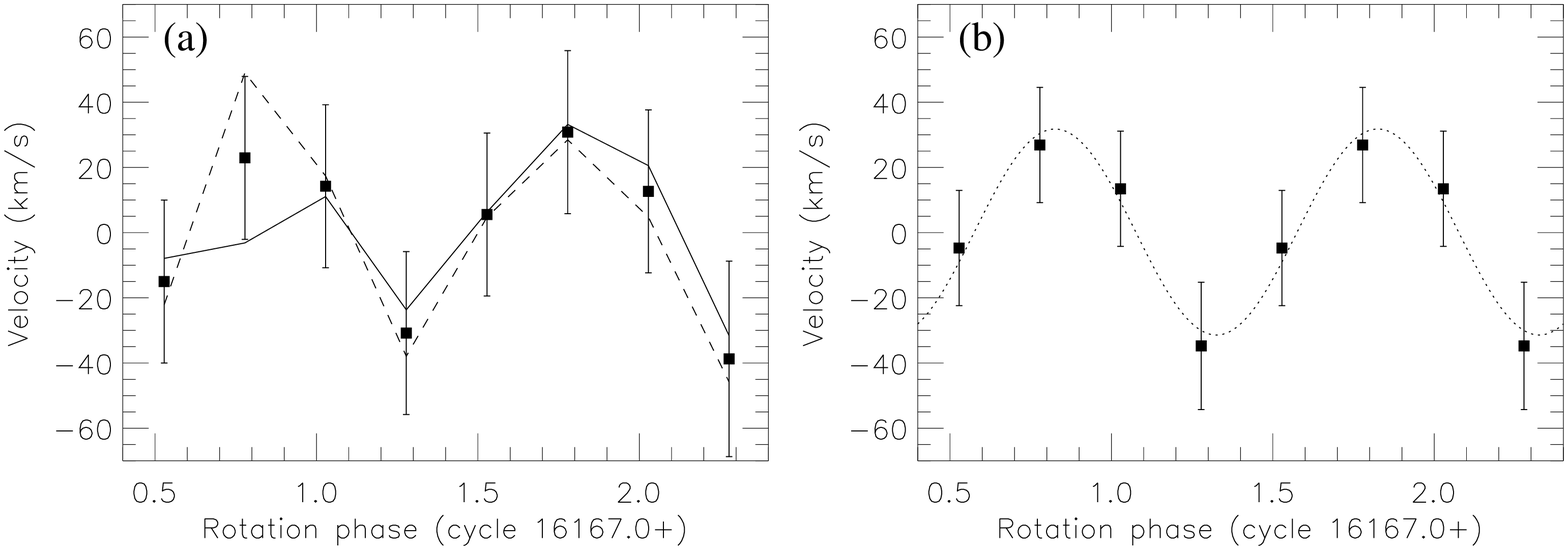}
\caption{ Velocity shifts in the line centroids of the O\,{\sc viii} 18.97\AA\ profile
($+$ is redshifted while $-$ is blue-shifted).
{\bf (a)}~Velocity shifts in the $+1$ and $-1$ orders (dashed and solid lines respectively)
are consistent. The squares represent the mean velocity variations
of the line profile from both orders and the error bars shown are computed using the errors from the
fitting routine in the  data analysis package {\em Sherpa}. 
Note that two consecutive phases were observed,  the rotation cycle number and phases 
are  calculated using the  ephemeris $H\!J\!D = 2\,444\,296.575 + 0.51479E$  (Innis et al.  1988).
{\bf (b)}~The phase-folded mean velocity shifts in the line centroids of the O{\sc viii} 18.97\,\AA\ profile ($+$ is redshifted while $-$ is blue-shifted). The dotted line is the best-fit sine-curve.
}
\label{fig:letgo8}
\end{figure}

The strongest line profiles are converted into velocity-space 
using the zero-velocity offsets described above. The $+1$ and $-1$ orders are 
analyzed separately as they have different offsets and  systematic distortions.
Centroids of the strongest lines in each phase-binned spectrum are also measured 
using the  Gaussian-fitting routine in {\em Sherpa} 
(as done with the zero-velocity calibrations). 
See Figure\,\ref{fig:letglins}a for an example
of  a  Gaussian fit to a phase-binned spectral line profile. 
Centroiding  is somewhat compromised in the short wavelength region despite 
having the highest counts and lowest background due to low spectral resolution.
Conversely, at longer wavelengths ($>88$\, \AA) the spectral resolution improves by a 
factor of more than 5 but centroiding is  difficult due to low counts and a relatively 
high background.

In this dataset, the strongest line  is the 
O\,{\sc viii} 18.97~\AA\ Ly$\alpha$ resonance line 
(approximately 260 counts total in each phase-binned O\,{\sc viii}
 line profile; see Figure\,1). 
Both the $+1$ and $-1$ orders show remarkably consistent velocity shifts 
(Figure\,\ref{fig:letgo8}a)
despite lines in different orders
being subject to different initial offsets and distortions (Chung et al. 2004). 
The level of agreement between both orders indicates that most 
systematic effects  are averaged out 
and the modulation shown here is due to the star itself.  
Figure\,\ref{fig:letgo8}b shows the rotational modulation traced out by the mean of the
velocity shifts from both orders. This pattern clearly repeats from one rotation cycle to the next.
The largest velocity shift has an amplitude of approximately $40$\kmsec, 
a relatively small fraction of AB Dor's photospheric \vsini\ value (90\kmsec).
The optimum sine-curve fit to this velocity modulation has a semi-amplitude 
of $30 \pm 10$\kmsec. We observe no  significant variation in the line flux of the 
phase-binned O\,{\sc viii} line profiles. 
Unfortunately, the lightcurve constructed by integrating the O\,{\sc viii} 
line is too noisy and cannot be used to  measure any rotational modulation.

\begin{figure}
\epsscale{1.0}
\plotone{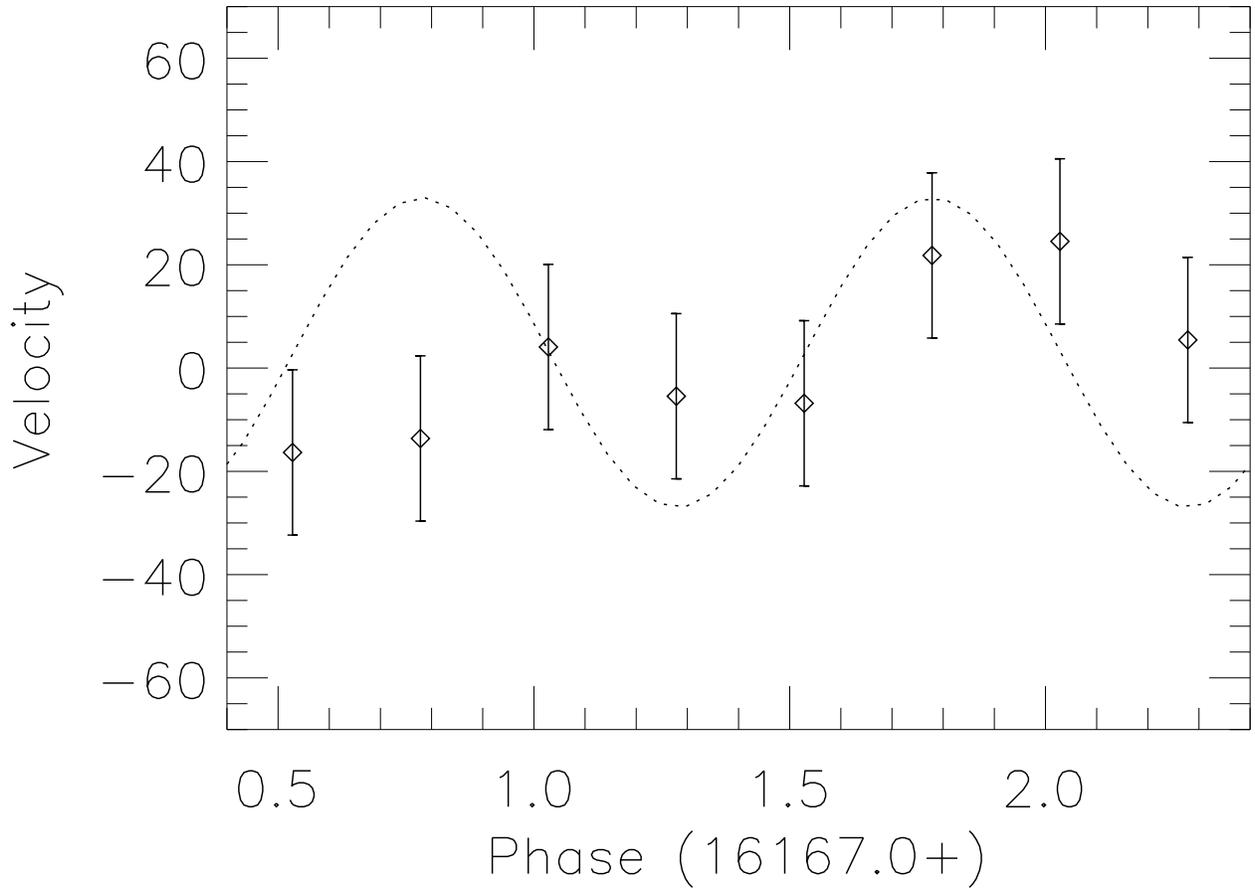}
\caption{ Velocity shifts in the line centroids of the  summed  line profile 
(diamonds). 
The error bars shown here are from Gaussian fitting only, but there are likely to be other
sources of error (see text for more details).
The rotation cycle and phase have been computed using the ephemeris of Innis et al.'s 
(1988; see Figure\,6b). The sine curve from Figure\,6b is also plotted here.
}
\label{fig:letgsum}
\end{figure}

\subsection{Centroid shifts at long wavelengths}

Since the centroiding accuracy of a line scales with the 
spectral resolution, the longer wavelength region of the spectrum should in 
principle be able to confirm the O\,{\sc viii} wavelength shifts.  Unfortunately, the 
count rate for individual lines is significantly less than for O\,{\sc viii}; 
however, there are a number of  sufficiently strong lines that can be used to 
produce a composite or summed line profile with a similar signal-to-noise ratio as the O{\sc viii} line, following the 
procedure described by Hoogerwerf, Brickhouse, \& Mauche (2004). Taking the 
quarter-phase binned spectra, we add up the signal  from the strongest
seven lines at wavelengths greater than 88\AA\ (see Table~1) as follows:
the  line profiles from both the $+1$ and $-1$ orders are first converted to velocity-space 
using  zero-velocity offsets  (computed from the total spectrum), 
they are then rebinned to the wavelength resolution (32\kmsec\ or 0.012\,\AA) corresponding to a line at the mean wavelength, 110\,\AA, and co-added.
Figure\,\ref{fig:letglins}b is an 
example of a summed line profile. 
As all the line profiles have slightly different velocity resolutions
the summed line profile will not be Gaussian and will be broader than the optimal 
single line profile.

The centroid position is determined for each quarter-phse summed 
line profile  by fitting Gaussian profiles, as with the O\,{\sc viii} line. This 
method produces centroids close to the flux-weighted mean center, but allows 
us to make a reasonable error estimate. Figure\,\ref{fig:letgsum} shows the velocity shifts with 
these errors in the summed profile as well as the sine-curve fit to the O\,{\sc viii} 
velocity shifts used in Figure\,\ref{fig:letgo8}. 
While the theoretical accuracy for a Gaussian line with the 
signal-to-noise ratio of the summed profile is about 5\,km\,s$^{-1}$ 
(see Hoogerwerf, Brickhouse \& Mauche 2004), the larger errors ($\sim$16\,km\,s$^{-1}$) 
appear to account for the non-Gaussian 
nature of the profile, as well as the high background. It should be noted, 
however,  that  there are systematic uncertainties in the centroids that may not be accounted 
for with this method. The ``zero velocity'' calibration for the individual 
lines in the total exposure is not as accurate as desired, particularly for 
the weaker lines, thus introducing additional profile effects. Furthermore, 
the wavelength distortions associated with the detector read-outs may affect 
certain measurements. Since each line profile samples two sets of readout 
taps during a dither cycle, two distinct profiles with different wavelength distortions are always being added. Differences in the 
amount of time spent on each side can lead to different profile shapes for the 
same line for different phases. 
Although we are only interested in {\it relative} differences, and the overall wavelength 
scale appears stable with time, mismatches may add an additional 10 to 15\,km\,s$^{-1}$ 
uncertainty, for the signal-to-noise of the summed profile, to what is derived from the 
Gaussian fits shown  in Figure\,\ref{fig:letgsum}. 
Improvements to the wavelength scale (Chung et al. 2005) 
may allow us to exploit the composite line technique 
with better accuracy in the future. In any case the individual  O\,{\sc viii} line 
measurements should be  robust.

\begin{deluxetable}{ccc}
\tablewidth{200pt}
\tablecaption{Thermal properties of  emission lines.}
\tablenum{1}
\tablehead{\colhead{Wavelength} & \colhead{Ion} & \colhead{Temperature} \\
\colhead{(\AA)} & \colhead{} & \colhead{(MK)}  } 
\startdata
 18.97 & O\,{\sc viii} & 3.16  \\
 88.08 & Ne\,{\sc viii}   & 0.631 \\
 93.92 & Fe\,{\sc xviii}  &  6.31 \\
 101.55 & Fe\,{\sc xix}   & 7.95 \\
 108.37 & Fe\,{\sc xix}   & 7.95 \\
 117.17 & Fe\,{\sc xxii}  & 12.6 \\
 128.73 & Fe\,{\sc xxi}   & 10.0 \\
 132.85 & Fe\,{\sc xxiii} & 12.6 \\
\enddata
\tablecomments{All lines except O\,{\sc viii} were used to generate summed line profiles.}
\end{deluxetable}

\section{Rotational broadening: LETG, HETG \& {\sc fuse} spectra}

\begin{figure}
\epsscale{1}
\plotone{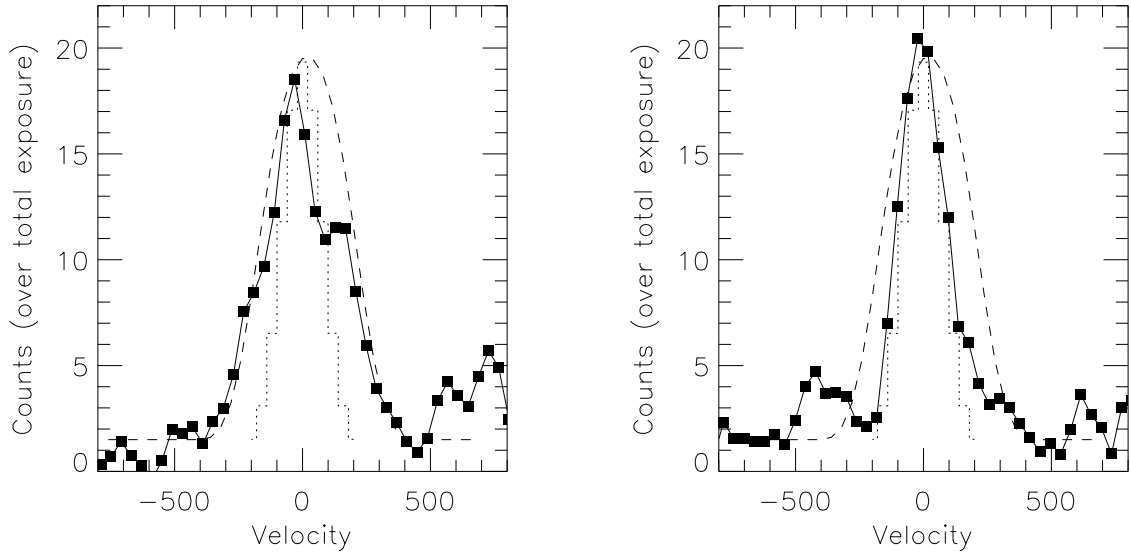}
\caption{The Fe\,{\sc xviii} 93.9\,\AA\ line profile from the LETG spectra, 
integrated over  88.1\,ksec. The $-1$ and $+1$ orders are plotted
left and right respectively. In each plot, the solid line and the filled symbols
show the observed data; the dotted line represents the thermally broadened
instrumental profile corresponding to a \vsini=90\kmsec\ (i.e. the photospheric
broadening); the dashed line corresponds to a \vsini=250\kmsec.
The $+1$ order profile is significantly narrower than the $-1$ order profile. This
is due to large uncertainties in the LETG dispersion calibration at 
long wavelengths.}
\label{fig:letgwidth}
\end{figure}

AB Dor's photospheric lines are substantially 
rotationally broadened (\vsini$=90$\,\kmsec). 
If the corona is a diffuse shell extended out to several \rstar,  LETG 
line profiles at long wavelengths
should be significantly broader than profiles subject to 
instrumental and thermal broadening effects alone.
While the spectral resolution is limited at short wavelengths, it improves to 
greater than 2000 above 90\,\AA\ and thus enables us to place
an upper limit on the extent of the X-ray emitting corona.
We measured the width of the strongest 
emission line profile in the long wavelength region, Fe\,{\sc xviii} 93.92\,\AA.
Unfortunately, as Figure\,\ref{fig:letgwidth} shows  there is little agreement between the 
$+1$ and $-1$ order line profiles;  when fitting the line profile width 
we find that the line profile widths are different by a factor of 2.
This is a further indication of the large uncertainties in the dispersion scale of the 
LETG/HRC-S setup at long wavelengths.

As the dispersion scale in the LETG line profiles is unreliable where the spectral
resolution is greatest, we measure line profile widths in X-ray and UV datasets of 
AB Dor obtained in previous years. While the corona of AB Dor may have altered
between epochs, these measurements enable us to place a reasonable limit on
the extent of the stellar corona.
We first look at the {\em Chandra} HETG/ACIS-S dataset acquired in 1999 October over 52\,ksec 
(Sanz-Forcada, Maggio \& Micela 2003). The spectra are extracted using standard procedures
except that pixel randomization is disabled (as with the LETG dataset).
In this dataset there are only a handful of strong emission lines that are suitable for 
this kind of analysis as only the high energy grating (HEG) has sufficient resolution 
to place a reliable upper limit on the line widths.  

\begin{figure}
\epsscale{1}
\plotone{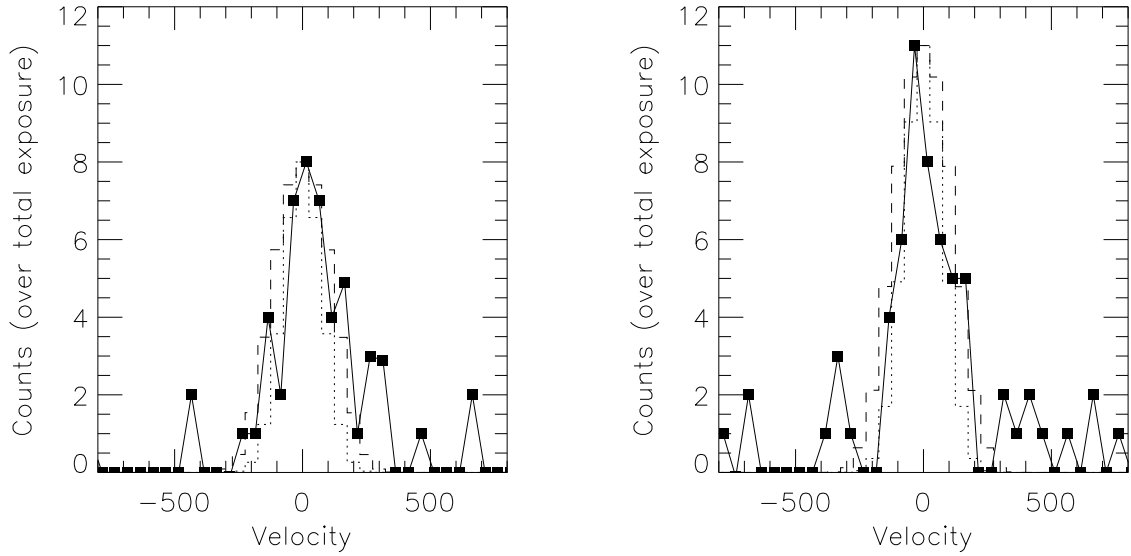}
\caption{The Fe\,{\sc xvii} 15\,\AA\ line profile from the HEG spectra 
of AB Dor taken in 1999, integrated over 52\,ksec. The $-1$ and $+1$ orders are plotted
left and right respectively. In each plot, the solid line and the filled symbols
show the observed data, the dotted line represents the thermally broadened
instrumental profile corresponding to a \vsini=90\,\kmsec\ (i.e. the photospheric value); 
the dashed line corresponds to a \vsini=158\,\kmsec\ ($\Delta R=0.75$\,\rstar).}
\label{fig:hetgwidth}
\end{figure}

In order to evaluate the amount of extra broadening in the line
profile after accounting for thermal and instrumental effects, 
we select the Fe\,{\sc xvii} 15.01\,\AA\ line profile as it has the least 
amount of thermal broadening amongst
the strongest emission lines in the HEG spectrum. 
We obtain the instrumental profile for that wavelength from the response 
matrix file, and convolve this with a 
Gaussian line profile corresponding to the
thermal broadening in the Fe\,{\sc xvii} line. This thermally broadened
instrumental profile is then convolved with 
rotation profiles corresponding to different \vsini\ values, representing
different coronal heights. Figure\,\ref{fig:hetgwidth} shows the 
observed line profiles 
from both the $+1$ and $-1$ orders, with line profiles broadened 
assuming \vsini\ values of 90\kmsec\ (dotted line) and 160\kmsec\ (dashed line):
corresponding to coronae extending 0.0\rstar\ and 0.75\rstar\ above
the stellar surface respectively. 
As the instrumental and thermal broadening dominate over
any small values of rotational broadening, we find 
that we can only place an upper limit on the extent of the X-ray corona. 
Our best fit to the spectra implies a corona extending $0.45 \pm 0.3$\rstar\ above the stellar
surface. Thus we place an upper limit of $\Delta R < 0.75$\rstar\ on the coronal extent. 

While this upper limit suggests that the bulk of the X-ray emitting corona is contained 
well within 1\rstar, its exact extent depends on its geometry, and how the coronal
loops are heated.  
The X-ray emission from coronal loops depends strongly on the orientation of
the loop on the stellar disk, the dominant temperature of individual coronal loops and 
the emissivity of the emission line (Alexander \& Katsev 1996, Wood \& Raymond 2000). 
The spectral line fitted here is integrated over a period lasting longer than 
1\prot\ and is thus a global average of all the coronal loops at a
range of viewing angles. 
It should also be noted that some X-ray emission may originate 
at greater heights but as it would  
contribute  predominantly to the wings of line profiles it would 
be difficult to detect and measure given the lack of counts. 
We will present detailed models of AB Dor's X-ray corona in a future paper 
based on extrapolations of photospheric magnetic field maps (acquired
using a simultaneous dataset) and different heating models. 

\begin{figure}
\epsscale{1.0}
\plotone{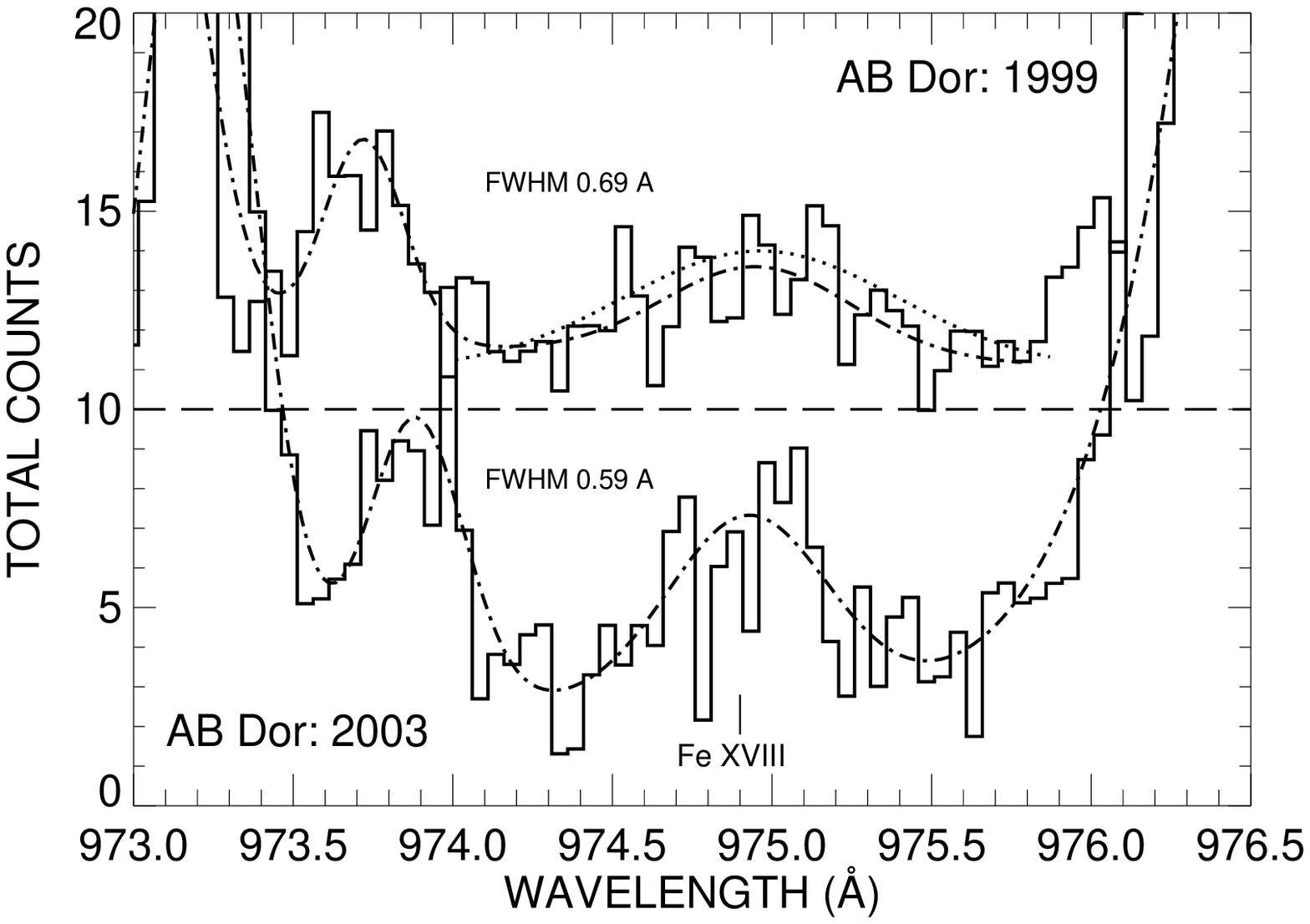}
\caption{ The {\sc fuse }  spectrum from the SiC2A channel
of the Fe\,{\sc xviii} line region in AB Dor. The lower spectrum
was observed in December 2003 (Dupree et al. 2005) with
a total exposure time of 108\,ks.
The upper spectrum (offset at a level of 10 counts)
obtained in 1999 (Redfield et al. 2003) had a total exposure time
of 24.2\,ks. Our multiple Gaussian fit is marked by the
broken line where the FWHM of the Fe\,{\sc xviii} line
is $\sim$ 0.6\,\AA.  A Gaussian curve with a FWHM of 1.0\,\AA\
(Redfield et al. 2003) is indicated by the dotted line
on the 1999 data.}
\label{fig:fusewidth}
\end{figure}

{\sc fuse} spectra of the coronal forbidden Fe\,{\sc xviii} ($\lambda$974) and Fe\,{\sc xix}
($\lambda$1118) lines have a
nominal spectral resolution of $\sim$ 20000 and so
the line profiles can be resolved. 
We have re-analyzed the coronal profiles
studied by Redfield et al. (2003), and also subjected our more recent
and longer {\sc fuse} observations of AB Dor to analysis (Dupree et al. 2005).  Because
the lines are weak, a fit to the count spectrum (as opposed
to the flux spectrum) is preferred in order to take proper
account of the Poisson error distribution and Cash statistics.
Moreover we have fit simultaneously the multiple nearby features as well 
as the coronal iron ions and continuum. Figure\,\ref{fig:fusewidth} shows the 
Fe\,{\sc xviii} transition
and the multiple Gaussian fit to the wing of the C\,{\sc iii} $\lambda$977
line, and the 2 O\,{\sc i} airglow lines.  The FWHM of the Fe\,{\sc xviii}
is 0.59$\pm$0.05\,\AA\ for the 2003 data. In addition a similar
procedure applied to the 1999 data suggests a Gaussian FWHM of 0.69$\pm$0.1\,\AA.
These fits are overplotted on the data in Figure\,{\ref{fig:fusewidth}} where
the broader line (FWHM =1.0\,\AA) suggested previously
(Redfield et al. 2003) is also shown.  Our multiple fitting
procedure gives consistent results of a FWHM $\sim$ 0.6\,\AA\
for both data sets. The FWHM value of 0.59\,\AA\
is to be preferred since the signal is stronger in the longer
exposure. The width of the Fe\,{\sc xviii} line is very sensitive to the
placement of the continuum and the presence of the broadened
line wing of C\,{\sc iii}. 
The Fe\,{\sc xix} line is more complicated  because it
is blended with many weak C\,{\sc i} lines and possibly an Fe\,{\sc ii}
line.  Nevertheless, the FWHM of the Fe XIX transition, 0.71$\pm$.05\,\AA,
is consistent with the value found for Fe\,{\sc xviii}.
Our measurements correspond to a coronal height, 
$\Delta R \approx 0.3\pm 0.2$\rstar.
As before, this value is an average of the corona over all observed phases. 

\section{Discussion}

The X-ray  lightcurve provides strong evidence for
rotational modulation in the corona of AB Dor.
The strongest peak in the power spectrum of this 
lightcurve is close to the star's rotation period 
of 0.51\,day. Sine-curve fitting reveals 
rotational modulation at the 12\% level, consistent
with previous estimates made using  ROSAT data.
The power  spectra also have strong peaks at periods 
corresponding to  0.32\,day and 0.18\,day. 
These  periods represent shorter 
term cyclic variations in AB Dor's corona caused by the presence
of three peaks in the lightcurve that are
evenly spaced in longitude and stable over  a period  lasting approximately
1.98 rotation cycles (i.e. the length of our exposure).
When studying the X-ray spectra we find that 
rotational modulation is seen as
Doppler shifts in the centroids of  phase-binned O{\sc viii} line profiles.
The centroid shifts trace a roughly sinusoidal pattern that repeats from one 
cycle to the next with a semi-amplitude of $\pm$30\kmsec\ (Figure\,{\ref{fig:letgo8}}b). 
As each line profile is integrated over 0.25\prot\ the maximum amplitude
variation of the centroid of the line profile is likely to be underestimated
(by about 11\%).
If only one compact active region is responsible for this modulation, 
it would be located at high latitudes, ($\approx 60$\degrees).
AB Dor's photospheric \vsini\ value is 91\kmsec, so 
emission from low latitudes would cause  Doppler shifts with larger amplitudes and 
emission from the poles  or from a diffuse X-ray emitting corona would not 
produce any Doppler shifts at all.
We can construct a simple model of the emitting corona by combining the diagnostics from the 
spectra and lightcurves presented in this paper.
Our models assume optically thin 
emission (Ness et al. 2003) and no limb brightening, based on 
line profiles of the FUSE O VI lines (Redfield et al. 2002).

\begin{figure}
\epsscale{1}
\plotone{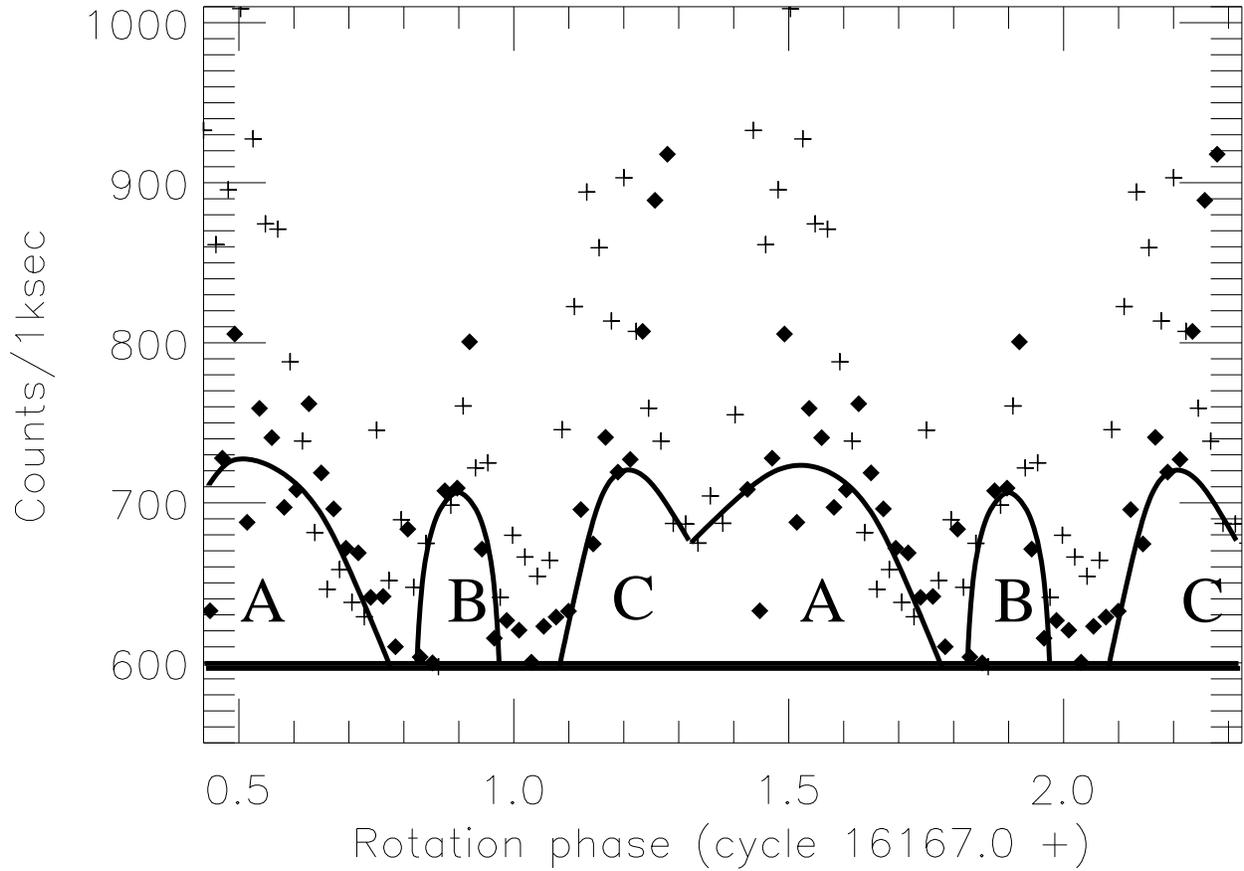}
\caption{ {\bf (a)}~{\em Chandra}/LETG lightcurve folded with AB Dor's rotation period.
Asterisks and diamonds represent consecutive rotation cycles.
The horizontal solid line represents the flat, unmodulated emission level while the
curves trace the quiescent modulated emission from the star.
}
\label{fig:discuss}
\end{figure}
 
As the base level in the lightcurve reveals, 
approximately 80\% of the emission from the star is  
unchanging. There are two possibilities to explain its lack of modulation:
either this component of the X-ray  emission originates at the pole of the star and/or 
this X-ray emission is homogeneously distributed, possibly   
extending to several stellar radii. However, our analysis of rotational broadening in 
{\sc fuse} spectra of AB Dor (albeit
from a different epoch) suggests that the bulk of the X-ray emitting corona, where
this  base level of strong emission would originate, is unlikely
to extend more than 0.5\rstar\ above the stellar surface.

The lightcurve also shows three peaks superimposed on the flat, plateau level of X-ray emission.  
These peaks show evidence of rotational modulation, 
although some also appear to be  intrinsically variable. The shorter-term variations
 (timescales $\ll 0.05$\prot) observed in 
the lightcurve peaks are likely to be caused by continuous flaring activity. 
However, we can estimate the quiescent modulation originating in these 
active regions by tracing the lower envelope of the variability in the lightcurve
(Figure\,\ref{fig:discuss}). Three peaks can be traced (regions A, B and C) in this lightcurve.
If we treat  each of these peaks as being caused by separate compact 
(radius $<5$\degrees) active regions, 
we can estimate their locations from the duration of the peaks and dips
in the lightcurve. Firstly, as the peaks only last a short time ($<0.5\,\prot$) this would
suggest that the active regions must be located close to the stellar surface with 
heights, $H\ll 0.5$\rstar. 
Secondly, the compact active regions are separated by over 100\degrees\ in
longitude from each other. As region B is only visible for 0.15\prot, 
it must be located in the partially obscured hemisphere of the star (AB Dor's inclination
angle is 60\degrees)  between $-55$\degrees\ and $-60$\degrees\ 
latitude with a height, $H \ll 0.5$\rstar.  
This region is unlikely to contribute to the centroid shifts in the O{\sc viii} line profiles 
as it  takes 0.15\prot\ to move from the approaching limb to the receding limb. 
As the phase-binned spectra are integrated over time-bins of 0.25\prot, the effective 
contribution to the centroid shift from this region is zero.  
Peak C in Figure\,\ref{fig:discuss} appears to last 0.22\prot\ and 
peak A lasts about 0.45\prot. If peaks A and C are caused by separate 
active regions, they are separated in longitude by 130\degrees  $\pm 15\degrees$.
Region C must then also be located in the partially obscured hemisphere
at a latitude between $-45$\degrees\ and $-60$\degrees\ latitude.
Region A can be a small region at equatorial latitudes (radius $<5$\degrees, latitude: 
$-5$\degrees\ $<\theta <5\degrees$) or a larger region at lower latitudes. 
The most likely explanation for the O{\sc viii} line centroid shifts is that 
they are  caused by a net effect due to the relative motions of different, stable
emitting regions in both hemispheres. Their exact contribution to the O{\sc viii} line centroids  
would be dependent on the temperatures of the active regions and their relative fluxes.  

\begin{deluxetable}{cccc}
\tablewidth{250pt}
\tablecaption{Properties of lightcurve peaks and inferred positions of active regions.}
\tablenum{2}
\tablehead{\colhead{Peak} & \colhead{Phase} & \colhead{Duration} & \colhead{Range of latitudes} \\
\colhead{} & \colhead{} & \colhead{\prot} & \colhead{} }
\startdata
 A       & 0.55 & 0.45 &  $-60$\degrees $ \le \theta \le 5$\degrees \\
 B       & 0.9   & 0.15 & $-60$\degrees $  \le \theta \le -55$\degrees  \\
 C       & 0.2   & 0.22 & $-60$\degrees $ \le \theta \le -45$\degrees \\
 A\,\&\,C & 0.35 & 0.65 & $-60$\degrees$\le \theta \le 40$\degrees \\
 \enddata
\tablecomments{Latitudinal positions are estimated assuming compact active regions 
(radius$\le 5$\degrees). }
\end{deluxetable}

Alternatively,  another scenario that would also fit the observations requires
peaks A and C in the lightcurve to arise from one active
region. The  dip in emission at phase 0.3 could be a shadowing effect
caused by cooler material transiting the visible disk at larger heights. 
Cool prominence-type complexes are commonly detected at heights
from 2\rstar\  to 6\rstar\ around AB Dor (e.g. Donati et al. 1999). 
These clouds have projected areas  of between 15\% to 20\% of the stellar disk 
as well as  absorbing column densities, 
$N_H \approx 10^{20}$\,cm$^{-2}$  
(Collier Cameron et al. 1990; Collier Cameron, Jardine \& Donati 200).
Evidence for soft X-ray emission being shadowed by cool circumstellar material has been 
observed on the Sun as well as in stars such as Proxima Centauri and the K dwarf 
component of the eclipsing binary V471\,Tauri (Haisch et al. 1983; 
Jensen et al. 1986; Walter 2004). 
The dip that we see near phase 0.3 requires absorbing column densities, 
$n_H \approx 10^{21}$\,cm$^{-2}$, which are an order of magnitude larger than those
estimated by Collier Cameron et al. (1990).  
Due to incomplete phase coverage of the second rotation phase
we are unable to ascertain whether or not this dip is a stable feature in both rotation cycles.
The active region in this scenario  would  be located near the surface, $H < 0.3$\rstar, 
at latitudes ranging from; $-60$\degrees\ to $40$\degrees, depending on its size.
If only one active region causes both  peaks  C and A as well as the centroid shifts, 
it should be observable near phase 0.1. Given the large integration times of the phase-binned
spectra, a compact active region at 40\degrees\ latitude might cause both the observed Doppler shifts
in the O{\sc viii} line profile as well as the lightcurve variability between phases 0.1 and 0.8.

\section{Conclusions}
To conclude, we construct a simple model of AB Dor's
X-ray emitting corona on the basis of the results discussed in this paper.
The observations we seek to fit using this simple model are listed below.

\begin{enumerate}

\item Rotationally modulated lightcurve: 
the lightcurve has a flat level of emission  superimposed with 
three peaks that cause 12\% rotational modulation.
There is also short-term variability that is probably caused by continuous flaring
in active regions.

\item Doppler shifts: The O{\sc viii} 18\,\AA\ line profile shows a rotationally 
modulated velocity pattern that repeats in two consecutive rotation cycles. The
best-fit sine curve has a semi-amplitude of 30 $\pm 10$\kmsec.

\item Rotational broadening: Assuming that Doppler broadening is the cause of
any excess broadening in the spectral line profiles (once thermal and instrumental effects
have been accounted for),  we find that 
the {\em Chandra}/HETG Fe{\sc xvii} 15\,\AA\ line profiles 
indicate that the emitting corona does not extend more than 
0.75\rstar\ above the stellar surface (\vsini$<  160$\kmsec). 
{\sc fuse} observations of the coronal forbidden Fe{\sc xviii} and Fe{\sc xix} 
lines at 974\,\AA\ and 
1118\,\AA\ respectively, suggest an even smaller upper limit of approximately 0.5\rstar 
(taking measurement errors into account). These limits assume a global average based on line
profiles that have been integrated for a period of more than one rotation period. 
\end{enumerate}

Assuming that the rotational modulation in the lightcurves and spectra are caused by
the presence of compact active regions, we construct a simple model of AB Dor's 
X-ray emitting corona at this epoch.  This model has 
an evenly distributed X-ray emitting corona extending less than 0.5\rstar\
 above the stellar surface. There  are  two to three compact active 
regions with heights, $H <0.3$\rstar. 
Assuming that all three active regions are compact (radii $<5\degrees$), two
of these regions are located
 in the partially obscured hemisphere of the (inclined) star at  latitudes ranging from
$-45$\degrees\ to  $-60$\degrees, while the third region may be located closer 
to the equator.

Alternatively, only one of the compact regions is located in the obscured hemisphere 
near a latitude,  $\theta \sim -55$\degrees, and another region is located at higher latitudes, 
near $40$\degrees. In this scenario, the higher latitude region
appears to be  shadowed by cool absorbing material near phase 0.3. 
It is likely that the unobscured hemisphere also contains more
compact emitting regions, but if they are evenly distributed in longitude they would not 
cause significant variability in the X-ray lightcurve.
However, they would still contribute to the  centroid shifts observed in the O{\sc viii} line profile.

Spectro-polarimetric ground-based observations of AB Dor were carried out
simultaneously with these {\em Chandra } observations. 
We will use Zeeman Doppler imaging techniques to map the surface magnetic field
of AB Dor at this epoch, and  will extrapolate these surface maps to produce 
detailed 3-D coronal magnetic field and X-ray emission models (e.g. Hussain et al. 2002, Jardine et al. 2002). 
In a future paper, we will use this X-ray emission model to predict rotational modulation, 
rotational broadening in coronal emission lines and 
compare these models to the X-ray observations presented here.

\acknowledgments

The authors received support through 
the Chandra guest observer grant, GO3-4022X.
GAJH was also funded through an ESA postdoctoral research fellowship.
NB acknowledges support from NASA contract NAS8-39073 to the
Smithsonian Astrophysical Observatory for the Chandra X-ray Center. 
The authors acknowledge the use of data analysis facilities provided by the Starlink Project which is run by CCLRC on behalf of PPARC. 
We would also like to thank the referee, Professor F. Walter for valuable comments which have improved the quality of this paper.

\end{document}